\documentclass[12pt]{article}
\pdfoutput=1
\usepackage{jhep-mod}
\usepackage{bm}
\usepackage{amssymb}
\title{Survey of analogue spacetimes}
\author{Matt Visser}
\affiliation{School of Mathematics, Statistics, and Operations Research, \\
Victoria University of Wellington, PO Box 600, Wellington 6140, New Zealand}
\emailAdd{matt.visser@msor.vuw.ac.nz}
\abstract{
Analogue spacetimes, (and more boldly, analogue models both of and for gravity), have attracted significant and increasing attention over the last decade and a half.
Perhaps the most straightforward physical example, which serves as a template for most of the others, is Bill Unruh's model for a dumb hole, (mute black hole, acoustic black hole), wherein sound is dragged along by a moving fluid --- and can even be trapped behind an acoustic horizon. This and related analogue models for curved spacetimes are useful in many ways: Analogue spacetimes provide general relativists with extremely concrete physical models to help focus their thinking, and conversely the techniques of curved spacetime can sometimes help improve our understanding of condensed matter and/or optical systems by providing an unexpected and countervailing viewpoint. In this chapter, I  shall provide a few simple examples of analogue spacetimes as general background for the rest of the contributions.   

\bigskip
\noindent
Submitted to the proceedings of the IX'th SIGRAV graduate school: \\
Analogue Gravity,  Lake Como, Italy, May 2011. 

\bigskip
\noindent
12 June 2012; \LaTeX-ed \today
}
\keywords{analogue spacetimes; analogue gravity; non-relativistic acoustics; surface gravity; relativistic acoustics; relativistic BECs; surface waves; blocking horizons.} 
\begin{document}
%---------------------------------------------------------------------------------------------------------------------------------------------
\maketitle
%------------------------------------------------------------------------------------------------------------------------------------------
% Basic definitions
%------------------------------------------------------------------------------------------------------------------------------------------
%------------------------------------------------------------------------------------------------------------------------------------------
%------------------------------------------------------------------------------------------------------------------------------------------
\def\R{{\mathbb{R}}}
\def\N{{\mathbb{N}}}
\def\Z{{\mathbb{Z}}}
\def\Q{{\mathbb{Q}}}
\def\d{{\mathrm{d}}}
\def\x{{\mathbf{x}}}
\def\n{{\mathbf{n}}}
\def\0{{\mathbf{0}}}
\def\v{{\mathbf{v}}}
\def\bnabla{{\mathbf{\nabla}}}
\def\widebar{\overline}
%------------------------------------------------------------------------------------------------------------------------------------------
%---------------------------------------------------------------------------------------------------------------------------------------------
\section{Introduction}
%---------------------------------------------------------------------------------------------------------------------------------------------
\label{S:introduction}
%---------------------------------------------------------------------------------------------------------------------------------------------
While the pre-history of analogue spacetimes is quite long and convoluted, with optics-based contributions dating as far back as the Gordon metric of 1923~\cite{Gordon}, significant attention from within the general relativity community dates back to Bill Unruh's PRL concerning acoustic black holes (dumb holes) published in 1981~\cite{Unruh:1980}. Even then, it is fair to say that the investigation of analogue spacetimes did not become mainstream until  the late 1990's. (See the recently updated Living Review article on ``Analogue gravity'' for a summary of the historical context~\cite{LRR}.)

In all of the analogue spacetimes, the key idea is to take some sort of ``excitation'', travelling on some sort of ``background'', and analyze its propagation in terms of the tools and methods of differential geometry. The first crucial technical distinction one has to make is between ``rays'' and ``waves''.
\begin{itemize}
\item 
The rays of ray optics (geometrical optics), ray acoustics (geometrical acoustics), or indeed any more general  ray-like phenomenon, are only concerned with the ``light cones'', ``sound cones'', or more generally the purely geometrical ``propagation cones'' defined by the ray propagation speed relative to the appropriate background. Physically, in this approximation one should think photons/phonons/ quasi-particles following some well-localized trajectory, rather than the more diffuse notion of a wave. Mathematically, we will soon see that it is appropriate to construct some metric $g_{ab}$, and some tangent vector $k^a$ to the particle trajectory,  such that:
\begin{equation}
g_{ab}\; k^a \; k^b = 0.
\end{equation}
Here indices such as $a$, $b$, $c$, \dots,  take on values in $\{0,1,2,3\}$, corresponding to both time and space, whereas indices such as $i$, $j$, $k$, \dots,  will take on values in $\{1,2,3\}$,  corresponding to space only. 
Of course we could multiply the metric by any scalar quantity without affecting this equation; this is known as a conformal transformation of the metric. (So distances change but angles are unaffected.) 
In the language of differential geometry, ray phenomena are sensitive only to a conformal class of Lorentzian geometries.
\item
In contrast, for waves one needs to write down some PDE --- some sort of wave equation. For example, for a scalar excitation $\Psi$ one needs to construct a wave equation in terms of a d'Alembertian~\cite{Unruh:1980, LRR, unexpected, Visser:1997, Visser:2001, carmen}:
\begin{equation}
{1\over\sqrt{-g}} \; \partial_a \left( \sqrt{-g} \; g^{ab} \; \partial_b \Psi\right) = 0.
\end{equation}
This d'Alembertian, (and in fact very many of the different possible types of wave equation), depends on all the components of the metric $g_{ab}$, not just the conformal class. (And conformal wave equations, of which the Maxwell electromagnetic wave equations are the most common, have their own somewhat different issues.)
\end{itemize}
In short, depending on exactly what one is trying to accomplish, one may sometimes be able to get away with ignoring an overall multiplicative conformal factor --- but for other applications knowledge of the conformal factor is utterly essential. 

What I shall now do is to present some elementary examples --- and a few not so elementary implications --- that will hopefully serve as a pedagogical  introduction to the more specific physics problems addressed in the other contributions to this volume.

\clearpage
%---------------------------------------------------------------------------------------------------------------------------------------------
\section{Optics: the Gordon metric and its generalizations}
%---------------------------------------------------------------------------------------------------------------------------------------------
\label{S:gordon}
%---------------------------------------------------------------------------------------------------------------------------------------------

The original Gordon metric~\cite{Gordon} of 1923  was limited to ray optics in a medium with a position-independent refractive index, and with some position-independent velocity.
Let 
\begin{equation}
\eta_{ab} = \left[\begin{array}{r|c}-1&0\\ \hline 0&\delta_{ij}\end{array} \right],
\end{equation}
%$\eta_{ab} = \mathrm{diag}\{-1,1,1,1\}$ 
denote (as usual) the special relativistic Minkowski metric, and correspondingly set the zeroth coordinate to $x^0 = t = c\,t_\mathrm{physical}$.  Denote the refractive index by $n$ and the 4-velocity of the medium by  $V^a = \gamma(1; \; \beta\, \n)$. Then we have  $V_a = \gamma(-1; \; \beta\, \n)$.  Now define
\begin{equation}
g_{ab} =  (\eta_{ab} + V_a V_b)   -  {V_a V_b\over n^2} = \eta_{ab} + \left(1-{1\over n^2}\right) V_a V_b. 
\end{equation}
In the rest frame of the medium $V^a\to(1;\;\0)$ and
\begin{equation}
g_{ab} \to \left[\begin{array}{c|c}-1/n^2&0\\ \hline 0&\delta_{ij}\end{array} \right].
\end{equation}
% $g_{ab} \to \mathrm{diag}\{-1/n^2,1,1,1\}$. 
 Therefore in this rest frame the null cones of the medium are exactly what we want:
\begin{equation}
0 = \d s^2 = -{\d t^2\over n^2} + \|\d \x\|^2 \qquad \implies \qquad \left\|{\d \x\over\d t}\right\| = {1\over n}.
\end{equation}
But more generally,  for non-zero velocity, $\beta\neq0$, the metric $g_{ab}$ provides a perfectly good special relativistic model for the light cones in a homogeneous moving medium. Let us agree to raise and lower the indices on the 4-velocity $V$ using the Minkowski metric $\eta$, then the contravariant Gordon metric is
\begin{equation}
g^{ab} =  (\eta^{ab} + V^a V^b)   -  n^2 \; V^a V^b. 
\end{equation}

The first and most obvious generalization is to note that one can easily make the refractive index and 4-velocity both space and time dependent. (Physically, this will certainly work as long as the wavelength and period of the light wavicle is short compared to the spatial and temporal scales over which changes of the background refractive index and 4-velocity are taking place.) 

A second generalization is to note that from the point of view of ray optics one might as well take
\begin{equation}
g_{ab} =  \Omega^2 \left[(\eta_{ab} + V_a V_b)   -  {V_a V_b\over n^2}\right]; 
\qquad
g^{ab} =  \Omega^{-2} \left[(\eta^{ab} + V^a V^b)   -  n^2 \; V^a V^b\right]. 
\end{equation}
The conformal factor $\Omega$ will simply drop out when determining the light cones. With the quantities $\Omega(x)$, $n(x)$, and $V(x)$ all being space and/or time dependent, this is the most general (but still physically natural) form of the (special relativity based) Gordon metric one can write. 

One can certainly calculate the Einstein tensor for this optical metric, but there is \emph{a priori} no really compelling reason to do so --- there is \emph{a priori} no good reason to attempt to enforce the Einstein equations for this optical metric, the physics is just completely different. (That being said, if one merely views this as an ansatz for interesting metrics to look at, then many of the standard spacetimes of general relativity can certainly be put into this form. For example, the Schwarzschild and Reissner--Nordstr\"om spacetimes, and the FLRW cosmologies,  can certainly be put in this form~\cite{Prado}.)

%---------------------------------------------------------------------------------------------------------------------------------------------
\paragraph{Example:}
%---------------------------------------------------------------------------------------------------------------------------------------------
Let us take $\Omega=n$ and write
\begin{equation}
g_{ab} =  n^2 (\eta_{ab} + V_a V_b)   -  V_a V_b. 
\end{equation}
This corresponds to 
\begin{equation}
\d s^2 = - \d t^2 + n^2 \; \|\d \x^2\|.
\end{equation}
Now pick the specific refractive-index profile
\begin{equation}
n =   {n_0\over 1 + r^2/a^2},
\end{equation}
so
\begin{equation}
\d s^2 = - \d t^2 + {n_0^2 \left[ \d r^2 + r^2\{ \d\theta^2+\sin^2\theta\;\d\phi^2\} \right]\over (1 + r^2/a^2)^2}.
\end{equation}
A theoretical cosmologist should recognize this as the Einstein static universe in isotropic coordinates~\cite{Tolman:1931, Robertson:1933, Tolman:1939}. 
A theoretician working in optics should recognize this as the Maxwell fish-eye lens~\cite{Maxwell1, Maxwell2, Maxwell3}. This is simply the first of many cross-connections between optics and general relativity. This becomes (or should become) a two-way street for information exchange.

The Maxwell fish-eye above is an example of a L\"uneburg lens~\cite{Luneburg}, and has now become the canonical example which helped initiate much of the recently developed field of ``transformation optics''~\cite{Leonhardt:2006,  Pendry:2006, Leonhardt:2006b,  Leonhardt:2008}. In particular, if one looks at this from the perspective of  a theoretical cosmologist then the prefect focussing properties are utterly trivial ---  after all, the spatial slices of the Einstein static universe are just the hyper-sphere $S^3$ in suitable coordinates --- the geodesics are obviously just great circles, which by symmetry must meet at the antipodes of the emission event, and so perfect focussing in the ray optics approximation is trivial. 

%---------------------------------------------------------------------------------------------------------------------------------------------
\paragraph{Limitations:}
%---------------------------------------------------------------------------------------------------------------------------------------------
Perhaps the greatest limitation of the Gordon metric is its inability (in its original 1923 formulation) to deal with wave properties of light. There is a rather non-trivial generalization to the full Maxwell equations~\cite{LRR},  but for technical reasons the generalization requires the very specific constraint
\begin{equation}
\hbox{[magnetic permittivity]} \propto \hbox{[electric permeability]}.
\end{equation}
For ordinary physical media this constraint is somewhat unphysical~\cite{LRR}, but there is hope that suitably designed metamaterials~\cite{metamaterials} may be designed to at least approximately satisfy this constraint. 

%---------------------------------------------------------------------------------------------------------------------------------------------
\paragraph{Foreground-background version:}
%---------------------------------------------------------------------------------------------------------------------------------------------
So far, the Gordon metric has been based on a optical medium in Minkowski space described by the flat metric $\eta_{ab}$. But now suppose we have a non-trivial background metric $f_{ab}$  arising from standard general relativity, and place a flowing optical medium on top of that. It now becomes interesting to consider the generalized Gordon metric
\begin{equation}
g_{ab} =  \Omega^2 \left[(f_{ab} + V_a V_b)   -  {V_a V_b\over n^2}\right];
\qquad
g^{ab} =  \Omega^{-2} \left[ (f^{ab} + V^a V^b)   -  n^2 \; V^a V^b \right]. 
\end{equation}
We now have the possibility of a non-trivial general relativistic background $f_{ab}(x)$, a position-dependent refractive index $n(x)$, a position-dependent 4-velocity $V^a(x)$, and a position-dependent conformal factor $\Omega(x)$. Note that the 4-velocity $V^a(x)$ now has to be a timelike unit vector with respect to the background metric $f_{ab}(x)$, and the indices on $V$ are raised and lowered using $f$.
In particular, note $g_{ab} V^a V^b = -\Omega^2/n^2$ and $g^{ab} V_a V_b = - n^2 /\Omega^2$, so it makes sense to define
\begin{equation}
\tilde V^a = {n\over \Omega} \; V^a; \qquad \hbox{and} \qquad \tilde V_a = {\Omega\over n}\; V_a. 
\end{equation}
Then $\tilde V$ is a timelike unit vector with respect to $g$, and its indices should be raised and lowered using $g$. Then we can adapt the generalized Gordon metric to also write the background $f$ in terms of the foreground $g$ as:
\begin{equation}
f_{ab} =  \Omega^{-2} \left[(g_{ab} + \tilde V_a \tilde V_b)   - n^2  \tilde V_a \tilde V_b\right];
\qquad
f^{ab} =  \Omega^{2} \left[ (g^{ab} + \tilde V^a \tilde V^b)   -  {\tilde V^a \tilde V^b\over n^2} \right]. 
\end{equation}
For a relativist the generalized Gordon metric provides an interesting ansatz for a potentially intriguing class of spacetimes to consider. From a theoretical optics perspective, one might view this procedure as an extremely general way of ``composing" and/or  ``inverting'' the transformations of transformation optics --- for example, one might first design some metamaterial~\cite{metamaterials} to generate the background $f_{ab}(x)$, and then impose some flowing optical medium on top of that. Various interesting possibilities come to mind. 

%---------------------------------------------------------------------------------------------------------------------------------------------
\section{Non-relativistic acoustics: the Unruh metric}
%---------------------------------------------------------------------------------------------------------------------------------------------
\label{S:non-relativistic}
%---------------------------------------------------------------------------------------------------------------------------------------------
Bill Unruh's 1981 PRL article~\cite{Unruh:1980}, and much of the follow up work~\cite{LRR, unexpected, Visser:1997},  was explicitly and intrinsically based on non-relativistic acoustics.  Let us explore the basic features of this particular model.
%---------------------------------------------------------------------------------------------------------------------------------------------
\paragraph{Geometric  acoustics:}
%---------------------------------------------------------------------------------------------------------------------------------------------
From the acoustic ray perspective the derivation is trivial: Let $c_s$ be the speed of sound, and let $\v$ be the velocity of the fluid. Then sound rays (phonon trajectories) satisfy~\cite{unexpected, Visser:1997, LRR}
\begin{equation}
\left\|\d\x - \v \; \d t\right\| = c_s \; \d t.
\end{equation}
Let us, already anticipating the possibility of an arbitrary conformal factor,  define
\begin{equation}
\d s^2 =  \Omega^2\left\{ - c_s \; \d t^2 + \left(\d\x - \v \; \d t\right)^2\right\}  = \Omega^2 \left\{ -(c_s^2-v^2) - 2\, \v \cdot \d\x \; dt + \|\d\x\|^2 \right\}. 
\end{equation}
(The zeroth coordinate is now most naturally chosen to simply be $x^0=t= t_\mathrm{physical}$, without any explicit factor of $c$. The speed of sound $c_s$ has the dimensions of a physical velocity.)
Then the sound-ray condition is completely equivalent geometrically to the  null-cone condition $\d s^2=0$.  In terms of a $4\times4$ matrix this is equivalent to defining the metric tensor~\cite{unexpected, Visser:1997, LRR}
\begin{equation}
g_{ab} = \Omega^2 \left[\begin{array}{c|c} -(c_s^2-v^2) & -v_j \\   \hline -v_i & \delta_{ij} \end{array} \right].
\end{equation}
The corresponding inverse metric is 
\begin{equation}
g^{ab} = \Omega^{-2} \left[\begin{array}{c|c} -1/c_s^2 & -v^j/c_s^2 \\   \hline -v^i/c_s^2 & \delta^{ij} - v^i v^j / c_s^2\end{array} \right].
\end{equation}
It should be emphasized that in this situation the velocity $\v$ and speed of sound $c_s$ will be inter-related in some (often quite complicated) manner --- the background fluid flow must satisfy the Euler equation and the continuity equation~\cite{Unruh:1980, unexpected, Visser:1997, LRR}.  

It should again further be emphasized (\emph{forcefully}) that while one can certainly calculate the Einstein tensor for this acoustic metric, there is \emph{a priori} no really compelling reason to do so --- there is \emph{a priori} no good reason to attempt to enforce the Einstein equations for this acoustic metric, the physics is just completely different. That being said, if one again views this as an ansatz for interesting metrics to look at, then many of the standard spacetimes of general relativity (but certainly not all interesting spacetimes) can be put into this form.
(For instance the Schwarzschild and Reissner--Nordstr\"om spacetimes can be put into this form by going to Painlev\'e--Gullstrand coordinates, but the Kerr and Kerr--Newman spacetimes \emph{cannot} be put in this form~\cite{LRR, Kerr-slice}.) 

To further develop the discussion, let us now introduce quantities
\begin{equation}
Q^{ab} = \left[\begin{array}{c|c} 0& 0 \\   \hline 0 & \delta^{ij}\end{array} \right];  \qquad V^a = (1;\; v^i) = (1;\;\v).
\end{equation}
Here the 4-velocity $V^a$ is normalized non-relativistically --- with the time component being unity.  Then for the inverse metric 
\begin{equation}
g^{ab} = \Omega^{-2} \left[Q^{ab} - {V^a V^b\over c_s^2}\right].
\end{equation}
But what about the covariant metric $g_{ab}$? Let us now define 
\begin{equation}
Q^\flat _{ab} = \left[\begin{array}{c|c} 0& 0 \\   \hline 0 & \delta_{ij}\end{array} \right].
\end{equation}
Then $Q^\flat _{ab}$ is the Moore--Penrose pseudo-inverse of $Q^{ab}$, and the object 
\begin{equation}
P^a{}_b = Q^{ac} \; Q_{cb} = \left[\begin{array}{c|c} 0& 0 \\   \hline 0 & \delta^i{}_j\end{array} \right]
\end{equation}
is a projection operator onto spatial slices.
Let us now furthermore define the quantities $V^\flat_a=Q^\flat_{ab} V^b = (0;\;v^i) = (0;\;\v)$, while $T_a=(1;\;\0)$. 
The best we can do for the covariant metric $g_{ab}$ is to now write the somewhat clumsy expression:
\begin{equation}
g_{ab} = \Omega^2\left[ Q^\flat_{ab} - (c_s^2-v^2) \,T_a \, T_b - T_a \,V^\flat_b - V^\flat_a \,T_b \right]. 
\end{equation}
In view of the fact that, with these definitions, one has $T_a \,V^a = 1$ and  $V^\flat_a \; V^a = v^2$, while $Q^{ab} \, T_b = 0$ and $Q^{ab} \, V^\flat_b =  V^a - T^a$, it is easy to verify that (as required)
\begin{equation}
g^{ab} \; g_{bc} = \delta^a{}_c.
\end{equation}
As we shall soon see, relativistic acoustics  is in some sense actually somewhat simpler than the non-relativistic case.   

%---------------------------------------------------------------------------------------------------------------------------------------------
\paragraph{Wave acoustics:}
%---------------------------------------------------------------------------------------------------------------------------------------------
If one goes beyond ray acoustics, then the parameter $\Omega$ is no longer arbitrary. One does have to make some additional (and rather stringent) technical assumptions --- barotropic, irrotational,  and inviscd (zero viscosity) flow~\cite{unexpected, Visser:1997, LRR}. Under those assumptions, by linearizing the Euler equation and continuity equation, after a little work one ultimately obtains a wave equation (a curved-spacetime d'Alembertian equation)  for perturbations of the velocity potential specified in terms of the density of the fluid and the speed of sound --- specifically one has $\Omega= \sqrt{\rho/c_s}$ in 3 space dimensions, $\Omega= \rho/c_s$ in 2 space dimensions, and technical problems arise in 1 space dimension.  Generally, in $d$ space dimensions, $\Omega = (\rho/c_s)^{1/(d-1)}$.  (See for example reference~\cite{LRR}.)

A specific feature of physical (wave) acoustics, not probed in the geometrical acoustics limit, is the behaviour of quasi-normal modes~\cite{QNM1, QNM2}. 
Furthermore, if the flow is not irrotational, so one is dealing with both background vorticity and vorticity-bearing perturbations, then a considerably more complicated \emph{system} of wave equations can be written down~\cite{Stone}, but this system of PDEs has nowhere near as clean a geometrical interpretation as the irrotational case.

%---------------------------------------------------------------------------------------------------------------------------------------------
\section{Horizons and ergo-surfaces in non-relativistic acoustics}
%---------------------------------------------------------------------------------------------------------------------------------------------

One of the very nice features of non-relativistic acoustics is that it is very simple and straightforward to define horizons and ergo-surfaces~\cite{unexpected, Visser:1997, LRR}.
To define these concepts, it is sufficient to  work in the geometric acoustics limit; wave acoustics adds additional constraints not directly needed to define horizons and ergo-surfaces. Consider for simplicity a stationary (time independent) configuration. 
\begin{itemize}
\item Ergo-surfaces are defined by the condition $\|\v\| = c_s$.
\item Horizons are surfaces, located for definiteness at $f(\x)=0$, that are defined by the 3-dimensional spatial condition $\vec\nabla f \cdot \v = c_s \; \|\vec\nabla f\|$. 
\end{itemize}
So the ergo-surface bounds the region where one cannot stand still without generating a sonic boom, and corresponds to Mach one, ($M\equiv v/c_s=1$). In contrast, on a horizon the \emph{normal component} of the fluid velocity equals the speed of sound, thereby either trapping or anti-trapping the acoustic excitations.

%---------------------------------------------------------------------------------------------------------------------------------------------
\paragraph{Stationary \emph{versus} static:}
%---------------------------------------------------------------------------------------------------------------------------------------------

In general relativity the words ``stationary'' and ``static'' have precise technical meanings that may not be obvious to non-experts. So a few words of explanation are called for:
\begin{itemize}

\item Stationary: For all practical purposes this means ``time independent''. More precisely, mathematically there is a Killing vector (a symmetry of the system) which is timelike at spatial infinity. Physically there is a class of natural time coordinates (not quite unique) in which the metric is time-independent. In this coordinate system the Killing vector is naturally associated with invariance under time translations $t\to t + C$. 

\item Static:   For all practical purposes this means ``time independent and non-rotating''.
More precisely, mathematically there is a Killing vector (a symmetry of the system) which is both timelike at spatial infinity \emph{and} ``hypersurface orthogonal'', meaning there exist functions $\xi(x)$ and $\tau(x)$ such that $K^a = \xi(x) \, g^{ab} \, \partial_b \tau(x)$.  Physically there is then a unique natural time coordinate, (in fact $\tau$, which is unfortunately not necessarily ``laboratory time''), in which the metric is both time-independent \emph{and} block-diagonal.  That is, with vanishing time-space components $g_{ti}=0$, in these coordinates the metric block diagonalizes into $(time)\oplus(space)$. The existence of a coordinate system with vanishing time-space metric components is sufficient in general relativity to imply zero angular momentum for the spacetime, and absence of ``frame dragging'', hence the sobriquet ``non-rotating''.  

\end{itemize}
A word of warning: Just because one \emph{can} always choose a coordinate system to block diagonalize a static spacetime does not mean this is always a good idea. Coordinates in which static spacetimes are block diagonal will break down at any horizon that might be present in the spacetime. (For instance, Schwarzschild geometry in the usual coordinates.)  
Permitting coordinates for static spacetimes which retain the manifest time independence, but do not explicitly force block diagonalization of the metric,  has significant technical and physical advantages. 
For one thing, this is the most natural situation when one works with ``laboratory time'' and a time independent fluid flow. For another thing, once one allows off-diagonal elements for the metric one can easily construct ``horizon penetrating'' coordinates, which are well defined both at and across the horizon. (For instance, Schwarzschild geometry in Painlev\'e--Gullstrand or Eddington--Finklestein coordinates.)  In particular, the acoustic metric as given above (in terms of laboratory time, speed of sound,  and fluid velocities) is automatically in horizon-penetrating form, all the components of both the metric $g_{ab}$ and its inverse $g^{ab}$ remain finite as one crosses the horizon.  
Let us now see how these ideas are used in practice.

%---------------------------------------------------------------------------------------------------------------------------------------------
\paragraph{Static configurations:}
%---------------------------------------------------------------------------------------------------------------------------------------------
Suppose the background flow satisfies the integrability constraint
\begin{equation}
{\v\over c_s^2 - v^2} = \vec\nabla \Phi, 
\end{equation}
and then consider the new time coordinate $\tau = t +\Phi$. (Here $t$ is explicitly laboratory time, while $\tau$ is constructed for mathematical convenience rather than for direct physical purposes.) Note that this integrability condition implies (but is stronger than) the vanishing of local helicity
\begin{equation}
h \equiv \v \cdot (\vec\nabla \times \v) = 0.
\end{equation}
In terms of this new time coordinate
\begin{equation}
\d s^2 = \Omega^2 \left\{ -(c_s^2-v^2)\; \d\tau^2 + \left[ \delta_{ij} + {v_i v_j \over c_s^2 - v^2} \right] \d x^i \d x^j  \right\}.
\end{equation}
The geometry is now in this form block-diagonal so it is \emph{manifestly} static, not just stationary. (And so the ergo-surfaces and horizons will automatically coincide.) The time translation Killing vector is 
\begin{equation}
K^a=(1;\;\0), \qquad \hbox{so} \qquad K_a=-\Omega^2(c_s^2-v^2;\;\0).
\end{equation}
To explicitly verify that this is hypersurface orthogonal in the sense defined above, note
\begin{equation}
K_a=-\Omega^2(c_s^2-v^2) \; \partial_a \tau,  \qquad \hbox{so}  \qquad  K^a = -\Omega^2(c_s^2-v^2) \; g^{ab} \; \partial_b \tau.
\end{equation}
The norm of this Killing vector is given by
\begin{equation}
K^a K_a = - \Omega^2 (c_s^2-v^2) = - \Omega^2\;c_s^2\;\left( 1 - {v^2\over c_s^2} \right).
\end{equation}
Using very standard techniques, the surface gravity is then calculable in terms of the gradient of this norm~\cite{LRR}. It is a standard result that for a Killing horizon the overall conformal factor drops out of the calculation~\cite{Kang}. Generalizing Unruh's original calculation~\cite{Unruh:1980}, which corresponds to $c_s$ being constant, one finds~\cite{unexpected, Visser:1997, LRR}
\begin{equation}
g_H =  {1\over2} \;\left\|\n\cdot\vec\nabla(c_s^2-v^2)\right\|_H = c_H \:\left\|\n\cdot\vec\nabla(c_s-v)\right\|_H = c_H \; \left| {\partial(c_s-v)\over\partial n}\right|_H,
\end{equation}
which can also be compactly written in terms of the Mach number $M \equiv v/c_s$ as
\begin{equation}
g_H = c_H^2\;  \left\|\n\cdot\vec\nabla(v/c_s)\right\|_H = c_H^2\;  \left\|\n\cdot\vec\nabla M\right\|_H = c_H^2 \; \left| {\partial M\over\partial n}\right|_H.
\end{equation}
We emphasise that this already works in the geometric acoustics framework, and that there is no need to make the more restrictive assumptions corresponding to wave acoustics that were made in references~\cite{LRR} and~\cite{Visser:1997}.
If the integrability condition is not satisfied one must be a little more devious.

%---------------------------------------------------------------------------------------------------------------------------------------------
\paragraph{Stationary but non-static configurations:}
%---------------------------------------------------------------------------------------------------------------------------------------------
If the acoustic horizon is stationary but not static there may or may not be additional symmetries, (in addition to the assumed time independence), so in particular the horizon may or may not be a Killing horizon. (A horizon is said to be a Killing horizon if and only if there exists \emph{some} Killing vector such that the location of the horizon coincides with the vanishing of the norm of that Killing vector.  So Killing horizons automatically satisfy nice symmetry properties.) For a Killing horizon the calculation of surface gravity is still relatively straightforward, for non-Killing horizons the situation is far more complex.

Note that in full generality, on the horizon we have $(\vec\nabla f \cdot \v)^2 = c_s^2 \; \|\vec\nabla f\|^2$, which we can rewrite in 3-dimensional form as $g^{ij} \, \partial_i f \, \partial_j f = 0$.  Since the configuration, and location of the horizon, is time independent this statement can be bootstrapped to 3+1 dimensions to see that \emph{on the horizon}
\begin{equation}
g^{ab} \; \nabla_a f \; \nabla_b f = 0. 
%,  \qquad \hbox{(on the horizon).}
\end{equation}
That is, the 4-vector $\nabla f$ is null on the horizon. In fact, on the horizon, where in terms of the 3-normal $\n$ we can decompose $\v_H = c_s \; \n +\v_\parallel$, we can furthermore write
\begin{equation}
(\nabla f)^a_H = \left(g^{ab} \; \nabla_b f\right)_H  =  {\|\vec\nabla f\|\over \Omega^2_H \; c_H} \; \left( 1; \v_\parallel\right)_H.
\end{equation}
That is, not only is the 4-vector $\nabla f$ null on the horizon, it is also a 4-tangent to the horizon --- so (as in general relativity) the horizon is ruled by a set of null curves. 
%These are even guaranteed to be geodesic null curves?
Furthermore, extending the 3-normal $\n$ to a region surrounding the horizon (for instance by taking $\n = \vec\nabla f/ \|\vec\nabla f\|$) we can quite generally write
$\v = v_\perp \; \n + \v_\parallel$.  Then away from the horizon
\begin{equation}
g^{ab} \; \nabla_a f \; \nabla_b f  =    { (c_s^2 - v_\perp^2)\;  \|\vec\nabla f\|^2  \over \Omega^2 \; c_s^2}.
\end{equation}
That is, the 4-vector $\nabla f$ is spacelike outside the horizon, null on the horizon, and timelike inside the horizon.  

%---------------------------------------------------------------------------------------------------------------------------------------------
\paragraph{Stationary but non-static Killing horizons:}
%---------------------------------------------------------------------------------------------------------------------------------------------
If the stationary horizon is Killing, then even if we do not explicitly know what the relevant Killing vector $\tilde K^a$ is, we know that its norm has to vanish on the horizon,  
and so the norm of this horizon-generating Killing vector is of the form
\begin{equation}
\tilde K^a \tilde K_a =  Q \;  (c_s^2-v_\perp^2) = - Q \;c_s^2\;\left( 1 - {v_\perp^2\over c_s^2} \right),
\end{equation}
for some unknown (but for current purposes irrelevant) function $Q$. Following closely the argument for the static case, \emph{mutatis mutandis}, we have~\cite{unexpected, Visser:1997, LRR}
\begin{equation}
g_H =  {1\over2} \;\left\|\n\cdot\vec\nabla(c_s^2-v_\perp^2)\right\|_H = c_H \:\left\|\n\cdot\vec\nabla(c_s-v_\perp)\right\|_H = c_H \; \left| {\partial(c_s-v_\perp)\over\partial n}\right|_H,
\end{equation}
which can also be compactly written in terms of the horizon-crossing Mach number, $M_\perp \equiv v_\perp/c_s$, as
\begin{equation}
g_H = c_H^2\;  \left\|\n\cdot\vec\nabla(v_\perp/c_s)\right\|_H = c_H^2\;  \left\|\n\cdot\vec\nabla M_\perp\right\|_H = c_H^2 \; \left| {\partial M_\perp\over\partial n}\right|_H.
\end{equation}
We again emphasise that this already works in the geometric acoustics framework, and that there is no need to make the more restrictive assumptions corresponding to wave acoustics that were made in references~\cite{LRR} and~\cite{Visser:1997}.
If the horizon is non-Killing then one must be even more devious.

%---------------------------------------------------------------------------------------------------------------------------------------------
\paragraph{Stationary but non-static non-Killing horizons:}
%---------------------------------------------------------------------------------------------------------------------------------------------
Such situations are, from a technical perspective, much more difficult to deal with. Such behaviour cannot occur in standard general relativity, where the Einstein equations stringently constrain the allowable spacetimes, but there seems no good reason to exclude it for acoustic horizons. Unfortunately, when it comes to explicit computations of the surface gravity there are still some unresolved technical issues.  
There is still a lot of opportunity for significant new physics hiding in these non-Killing horizons.

%---------------------------------------------------------------------------------------------------------------------------------------------
\section{Relativistic acoustics}
%---------------------------------------------------------------------------------------------------------------------------------------------
\label{S:relativistic}
%---------------------------------------------------------------------------------------------------------------------------------------------
Full relativistic acoustics (either special relativistic or general relativistic) adds a few other quirks which I briefly describe below. (See early astrophysical work by Moncrief~\cite{Moncrief}, a more recent cosmological framework developed in~\cite{Vikman}, and a pedagogical exposition in reference~\cite{carmen} for details.) Note that the interest in, and need for, relativistic acoustics is driven by astrophysical and cosmological considerations, not by direct laboratory applications. There are at least three situations in which relativistic acoustics is important:
\begin{itemize}

\item Speed of sound comparable to that of light.\\
In any ideal gas once $kT \gg m_0c^2$ then $p \approx {1\over3} \rho$ and so $c_s \approx {1\over\sqrt{3}} c$.   \\
This is physically relevant, for instance, in various stages of big bang cosmology.

\item Speed of fluid flow comparable to that of light.\\
This is physically relevant, for instance,  in some black hole accretion disks and/or the jets emerging from active galactic nuclei (AGNs).

\item Tight binding: $p \lesssim \varrho$ or $|\mu| \ll m_0 c^2$.\\
Once the pressure is an appreciable fraction of the energy density, or the absolute value of the chemical potential is much smaller than the rest mass, then the usual derivation of the conformal factor appearing in the wave version of the acoustic metric must be significantly modified.\\
This is physically relevant, for instance, in cores of neutron stars. 

\end{itemize}
It is somewhat unclear at present as to whether relativistic acoustics can be made directly relevant for laboratory physics. Some first steps in this regard may be found in reference~\cite{Fagnocchi:2010},  where the possibility of experimentally constructing relativistic BECs is considered.

%---------------------------------------------------------------------------------------------------------------------------------------------
\paragraph{Geometric  acoustics:}
%---------------------------------------------------------------------------------------------------------------------------------------------
If one works with special relativistic acoustics, rather than non-relativistic acoustics,  then at the level of ray acoustics one will simply obtain an acoustic variant of the Gordon optical metric
\begin{equation}
g_{ab} =  \Omega^2 \left[(\eta_{ab} + V_a V_b)   -   {c_s^2\over c^2} \; V_a V_b \right]. 
\end{equation}
The only difference is that the refractive index has now been replaced by the ratio of the speed of sound to the speed of light: $n^{-1}(x) \to c_s(x)/c$. (The 4-velocity of the medium is still $V^a(x)$,  and the conformal factor $\Omega(x)$ is still undetermined.) In general relativistic acoustics this would become
\begin{equation}
g_{ab} =  \Omega^2 \left[(f_{ab} + V_a V_b)   -   {c_s^2\over c^2} \; V_a V_b \right],
\end{equation}
where $f_{ab}(x)$ is now the general relativistic physical background metric obtained by solving the Einstein equations, and $g_{ab}$ is the acoustic metric for the acoustic perturbations in the fluid flow.   Note that the 4-velocity $V^a(x)$ now has to be a timelike unit vector with respect to the background metric $f_{ab}(x)$. For ray acoustics this is all one can say.

%---------------------------------------------------------------------------------------------------------------------------------------------
\paragraph{Wave  acoustics:}
%---------------------------------------------------------------------------------------------------------------------------------------------

One can again go to wave acoustics, deriving a wave equation by linearizing the general-relativistic version of the Euler equations. The same sort of technical assumptions must be made, (irrotational, barotropic, and inviscid),  and one now obtains a slightly more complicated formula for the conformal factor~\cite{carmen}
\begin{equation}
\Omega = \left( { n^2\over c_s(\varrho+p)} \right)^{1/(d-1)}\hskip-30pt.
\end{equation}
Here $n$ is the number density of particles, and $\varrho$ is the energy density (rather than the mass density $\rho$), while $c_s$ is the speed of sound. The quantity $p$ is the pressure,  and $d$ is the number of space dimensions.

%---------------------------------------------------------------------------------------------------------------------------------------------
\paragraph{Non-relativistic limit:}
%---------------------------------------------------------------------------------------------------------------------------------------------

In the non-relativistic limit among other things we certainly have $p \ll \varrho$. Also in terms of the average particle mass $\widebar m$ one has 
\begin{equation}
\varrho = \rho \, c^2 \approx n \,\widebar m \, c^2,
\end{equation}
and so
\begin{equation}
 { n^2\over c_s(\varrho+p)} \approx { n^2\over c_s\varrho} \approx  { n\over c_s \; (\widebar m c^2)} = { \rho \over c_s \; (\widebar m^2 c^2)}  \propto {\rho\over c_s},
\end{equation}
thereby (as required for internal consistency) reproducing the correct limit for the conformal factor. 

The correct limit for the tensor structure is more subtle. (A suitable discussion can be found in reference~\cite{carmen}.)
Formally taking the limit $c\to\infty$, but holding $c_s$ and $v$ fixed, a brief calculation yields:
\begin{equation}
g_{00} = \Omega^2\left[ -1 + \gamma^2 - {c_s^2\over c^2} \gamma^2 \right] \to - \Omega^2 \; {c_s^2- v^2\over c^2} + \dots
\end{equation}
\begin{equation}
g_{0i} = -\Omega^2\left[1 - {c_s^2\over c^2}\right] \gamma^2\beta_i \to  -\Omega^2 \; {v\over c} + \dots
\end{equation}
\begin{equation}
g_{0i} = \Omega^2\left\{ \delta_{ij}  -\left[1 - {c_s^2\over c^2}\right] \gamma^2\beta_i \beta_j \right\} \to \Omega^2 \; \delta_{ij} + \dots
\end{equation}
Then, switching from $(c\,t,\;\x)$ coordinates to $(t,\;\x)$ coordinates, the relativistic $g_{ab}$ of this section correctly limits to the non-relativistic $g_{ab}$ of the previous section.

%---------------------------------------------------------------------------------------------------------------------------------------------
\section{Bose--Einstein condensates}
%---------------------------------------------------------------------------------------------------------------------------------------------
\label{S:BECs}
%---------------------------------------------------------------------------------------------------------------------------------------------

Bose--Einstein condensates (BECs) provide a particularly interesting analogue model because they are relatively easy to construct and manipulate in the laboratory, and specifically because the speed of sound is as low as a few centimetres per second.  Most work along these lines has focussed on non-relativistic BECs. Suitable background references are~\cite{Garay:1999, Barcelo:2000, Barcelo:2001, Barcelo:2003, Lahav:2009, Jannes:2009, Mayoral:2010}. See also the companion chapter by Balbinot~\emph{et al.}~in the current volume~\cite{Balbinot:Como}.  In view of the coverage of this topic already provided in that chapter, I shall not have more to say about it here. 

In contrast, I will briefly discuss the relativistic BEC model of Fagnocchi \emph{et al.} that is presented in reference~\cite{Fagnocchi:2010}.  While relativistic BECs do not seem currently to be a realistic experimental possibility, the theoretical treatment introduces some new issues and effects. The relativistic BECs naturally lead to two quasiparticle excitations, one massless and one massive, with rather complicated excitation spectra. (In this sense the relativistic BECs are reminiscent of the ``massive phonon'' models obtained from multiple mutually interacting non-relativistic BECs~\cite{Visser:2005, Visser:2004, Liberati:2005}.)  In the relativistic BEC one obtains a 4th-order differential wave equation for the excitations, which is ultimately why one has two branches of quasiparticle excitations. In the limit where the relativistic generalization of the so-called quantum potential can be neglected, the wave equation simplifies to the d'Alembertian equation --- for a relativistic acoustic metric of the generalized Gordon form discussed in the previous section. In the limit where both relativistic effects and the quantum potential can be neglected, one recovers the (wave acoustic version of) Unruh's non-relativistic acoustic metric.

%---------------------------------------------------------------------------------------------------------------------------------------------
\paragraph{Madelung representation:}
%---------------------------------------------------------------------------------------------------------------------------------------------
There is an important and non-obvious technical point to be made regarding the linearization of the Madelung representation in a BEC context, (or in fact in any situation where one is daeling with a non-linear Schr\"odinger-like equation).  For any complex field $\psi$ the Madelung representation is
\begin{equation}
\psi = \sqrt{\rho} \; e^{i \phi}.
\end{equation}
When linearizing, (which is the basis of separating the system into background plus excitation, or condensate plus quasiparticle), there are at least three things one might envisage doing:
\begin{enumerate}
\item  Take $\psi = \psi_0 + \epsilon \; \psi_1 + \mathcal{O}(\epsilon^2)$.
\item  Take $\rho = \rho_0 + \epsilon \; \rho_1 + \mathcal{O}(\epsilon^2)$, and $\phi = \phi_0 + \epsilon \; \phi_1 + \mathcal{O}(\epsilon^2)$.
\item  Take $\psi = \psi_0 \left\{ 1+ \epsilon \; \chi + \mathcal{O}(\epsilon^2) \right\}$.
\end{enumerate}
Note that routes 2 and 3 are related by:
\begin{equation}
{\rho_1\over\rho_0} = {\chi+\chi^\dagger\over2}; \qquad \qquad  \phi_1 = {\chi-\chi^\dagger\over2i}. 
\end{equation}
Mathematically, all three routes must carry the same intrinsic physical information, but the clarity with which the information can be extracted varies widely depending on the manner in which the perturbative analysis is presented. When actually carrying out the linearization, it turns out that route 1 is never particularly useful, and that routes 2 and 3 are essentially equivalent for a non-relativistic BEC, ultimately leading to formally identical wave equations. In contrast, for relativistic BECs it is route 3 that leads to the cleanest representation~\cite{Fagnocchi:2010}, while route 2 leads to a bit of a mess~\cite{LRR}.  (A mess involving integro-differential equations.) This is not supposed to be obvious, and will not be obvious unless one tries to carefully work through the relevant technical literature. With hindsight, route 3 appears to be the superior way of organizing the perturbative calculation.

%---------------------------------------------------------------------------------------------------------------------------------------------
\section{Surface waves and blocking horizons}
%---------------------------------------------------------------------------------------------------------------------------------------------
\label{S:surface}
%---------------------------------------------------------------------------------------------------------------------------------------------

Surface waves (water-air, or more generally waves on any fluid-fluid interface) are described by an incredibly complex and subtle theoretical framework --- one of the major technical complications comes from the fact that surface waves are highly dispersive, with a propagation speed that is very strongly frequency dependent. Thus, insofar as one can put surface wave propagation into a Lorentzian metric framework, one will have to adopt a ``rainbow metric'' formalism with a frequency dependent metric. The trade-off is that this system is relatively easily amenable to laboratory investigation through ``wave tank'' technology~\cite{Badulin, Silke&Bill, Rousseaux:2007}. 

%---------------------------------------------------------------------------------------------------------------------------------------------
\paragraph{Surface waves in the geometric limit:}
%---------------------------------------------------------------------------------------------------------------------------------------------
As long as the wavelength and period of the surface wave are small compared to the distances and timescales on which the depth of water is changing one can usefully work in the geometric (ray) limit.  Under those conditions one can write a 2+1 dimensional metric to describe ray propagation:
\begin{equation}
\d s^2 =  \Omega^2\left\{ - c_\mathrm{sw}^2 \; \d t^2 + \left(\d\x - \v \; \d t\right)^2\right\}  = \Omega^2 \left\{ -(c_\mathrm{sw}^2-v^2) - 2\, \v \cdot \d\x \; dt + \|\d\x\|^2 \right\}. 
\end{equation}
Here $c_\mathrm{sw}$ is the speed of the surface waves in the comoving frame (that is, comoving with the surface of the fluid), and $\v$ is the (horizontal) velocity of the surface. Unfortunately the speed $c_\mathrm{sw}$ is a relatively complicated function of (comoving) frequency, depth of the water, density of the fluid, the acceleration due to gravity, the surface tension, \emph{etcetera}. (See for instance references~\cite{toy, pedagogical, Rousseaux:2010, Rousseaux:Como}.)

Now  in terms of a $3\times3$ matrix, this is equivalent to defining the metric tensor
\begin{equation}
g_{ab} = \Omega^2 \left[\begin{array}{c|c} -(c_\mathrm{sw}^2-v^2) & -v_j \\   \hline -v_i & \delta_{ij} \end{array} \right].
\end{equation}
The indices  $a$, $b$, $c$, \dots, take on values in $\{0,1,2\}$, corresponding to both time and (horizontal) space, whereas indices such as $i$, $j$, $k$, \dots,  take on values in $\{1,2\}$,  corresponding to (horizontal) space only. 
The corresponding inverse metric is 
\begin{equation}
g^{ab} = \Omega^{-2} \left[\begin{array}{c|c} -1/c_\mathrm{sw}^2 & -v^j/c_\mathrm{sw}^2 \\   \hline -v^i/c_\mathrm{sw}^2 & \delta^{ij} - v^i v^j /c_\mathrm{sw}^2\end{array} \right].
\end{equation}
In the fluid dynamics community, one most often restricts attention to 1 spatial dimension, then surface waves are said to be ``blocked'' whenever one has $\|\v\| > c_\mathrm{sw}$, and one will encounter considerable attention paid to this concept of ``wave blocking'' in that community.  This is what a general relativist would instead call ``trapping'', and consequently the mixed terminology phrase ``blocking horizon'' has now come into use within the analogue spacetime community. Note that instead of speaking of Mach number (appropriate for acoustic propagation through the bulk of a medium),  in a surface wave context it is the Froude number that governs the formation of ergo-regions and horizons.

%---------------------------------------------------------------------------------------------------------------------------------------------
\paragraph{Surface waves in the physical limit:}
%---------------------------------------------------------------------------------------------------------------------------------------------
Moving beyond the geometric/ray approximation for surface waves is mathematically rather tricky. Within the fluid dynamics community relevant work is based on the Boussinesq approximation~\cite{Boussinesq:1871, Boussinesq:1872}, and its modern variants~\cite{Madsen}. Within the analogue spacetime community, see particularly the basic theoretical work in reference~\cite{Schutzhold:2002}, and in the related chapter~\cite{Unruh:Como} in this volume.  (See also~\cite{Chaline:Como, Carusotto:Como} for a more applied perspective.) Physically, in addition to the presence of dispersion,  a second complicating issue is this:  The fluid at the surface is moving both vertically (the wave) and horizontally (the background flow), while at the base of the fluid (which may be at variable depth), the no-slip boundary condition enforces zero velocity. 

Based on the three-dimensional Euler and continuity equations one then has to construct an interpolating model for the fluid flow that connects the surface motion to the zero-velocity  motion at the (variable depth) base. Once this is achieved, one throws away the interpolating model and concentrates only on the physical observable: the motion of the surface. The analysis is mathematically and physically subtle, and (in the physical or wave limit) the theoretical framework for surface waves is nowhere near as clean and straightforward as for barotropic inviscid irrotational acoustic perturbations travelling through the bulk.

%---------------------------------------------------------------------------------------------------------------------------------------------
\paragraph{Experiments:}
%---------------------------------------------------------------------------------------------------------------------------------------------
The key benefit of surface waves is that the propagation speed $c_\mathrm{sw}$ is easily controllable by adjusting the depth of fluid, that background flows are easily set up by simple mechanical pumps, and that ``wave tank'' and related technologies are well understood and well developed.  (See for example, the early 1983 experiments by Badulin \emph{et al.}~\cite{Badulin}.) This particular analogue spacetime has recently led to several very interesting experimental efforts~\cite{Silke&Bill, Rousseaux:2007, Jannes:2010}. For instance, Weinfurtner \emph{et al.}~have performed an experiment looking at  the classical (stimulated) analogue of Hawking radiation from a blocking horizon, and have detected an approximately Boltzmann spectrum of Hawking-like modes~\cite{Silke&Bill}, while Rousseaux~\emph{et al.} have experimentally investigated the related ``negative-norm modes''~\cite{Rousseaux:2007}. The relation between the ``hydraulic jump'' and blocking horizons has been experimentally investigated by Jannes~\emph{et al.}~\cite{Jannes:2010}. Some related theoretical developments are reported in~\cite{Rousseaux:2010, Volovik:2005}. Work on this topic is ongoing.

%---------------------------------------------------------------------------------------------------------------------------------------------
\section{Optical fibres/optical glass}
%---------------------------------------------------------------------------------------------------------------------------------------------
\label{S:fibres}
%---------------------------------------------------------------------------------------------------------------------------------------------

In an optical context, related ``optical blocking'' phenomena occur when a ``refractive index pulse'' (RIP) is initiated in an optical fibre~\cite{fibre}, or in optical glass~\cite{glass1, glass2, glass3}. The basic idea is that things are arranged so that while the RIP moves at some speed $v_\mathrm{RIP}$, the velocity of light outside the RIP is greater than the velocity of the RIP $c_\mathrm{outside} > v_\mathrm{RIP}$,  while  inside the RIP we have the contrary situation $c_\mathrm{inside} < v_\mathrm{RIP}$. This, (certainly within the geometric optics framework), sets up a ``black'' horizon at the leading edge of the RIP, and a ``white'' horizon at the trailing edge. (For technical details see references~\cite{fibre, glass1, glass2, glass3}.) Some subtleties of the theoretical analysis lie in the distinction between group and phase velocities --- are we dealing with ``phase velocity horizons'' or ``group velocity horizons''? Other technical subtleties have to do with the transition from geometric optics to wave optics --- there are a number of complex and messy technical details involved in this step. 

An intriguing experiment  based on these ideas has been carried out by Belgiorno~\emph{et al.}, with results reported in reference~\cite{Belgiorno:2010}.  While it is clear that some form of quantum radiation has been detected, there is still some disagreement as to whether this is (analogue) Hawking radiation, or possibly some other form of quantum vacuum radiation~\cite{Schutzhold:2010, Belgiorno:2010b, Prain}. Work on this topic is ongoing. 

%---------------------------------------------------------------------------------------------------------------------------------------------
\section{Other models}
%---------------------------------------------------------------------------------------------------------------------------------------------
\label{S:other}
%---------------------------------------------------------------------------------------------------------------------------------------------

A complete and exhaustive catalogue of other analogue models would be impractical. See the Living Review article on ``Analogue gravity'' for more details~\cite{LRR}. Selected models, (a necessarily incomplete list), include:
\begin{itemize}
\item Electromagnetic wave guides~\cite{pulse:2005}. 
\item Graphene~\cite{Cortijo:2006, Vozmediano:2010}.
\item Slow light~\cite{Leonhardt:2000, Visser:2000, Leonhardt:2000b, Leonhardt:2000c, Leonhardt:2001}.
\item Liquid helium~\cite{Volovik:2000, Jacobson:1998}.
\item Fermi gasses~\cite{Giovanazzi:2004, Giovanazzi:2011}.
\item Ion rings~\cite{Horstmann:2009}.
\end{itemize}
Beyond the issue of simply \emph{developing} analogue models, there is the whole subject of \emph{using} analogue models to probe, (either theoretically or more boldly experimentally), a whole raft of physics questions such as directly verifying the existence of Hawking radiation, the possibility of Lorentz symmetry violations~\cite{Liberati:2012}, the nature of the quantum vacuum~\cite{Volovik:2011kg, Jannes:2011em, Finazzi:2012wz}, \emph{etcetera.} For more details, see reference~\cite{LRR}, and other chapters in this volume. 

%---------------------------------------------------------------------------------------------------------------------------------------------
\section{Discussion}
%---------------------------------------------------------------------------------------------------------------------------------------------
\label{S:discussion}
%---------------------------------------------------------------------------------------------------------------------------------------------

The general theme to be extracted from these considerations is this: The propagation of excitations (either particles or waves) over a background can often (not always) be given a geometric interpretation in therms of some ``analogue spacetime''. As such a geometric interpretation exists, there is a strong likelihood of significant cross-fertilization of ideas and techniques between general relativity and other branches of physics. Such possibilities have increasingly attracted attention over the last decade, for many reasons. The other chapters in these proceedings will explore these ideas in more specific detail.

%---------------------------------------------------------------------------------------------------------------------------------------------
\section*{Acknowledgments} 
%---------------------------------------------------------------------------------------------------------------------------------------------

This research was supported by the Marsden Fund, and by a James Cook Research Fellowship, both 
administered by the Royal Society of New Zealand.  

%---------------------------------------------------------------------------------------------------------------------------------------------

%---------------------------------------------------------------------------------------------------------------------------------------------

\begin{thebibliography}{99}
%---------------------------------------------------------------------------------------------------------------------------------------------

\bibitem{Gordon}
W.~Gordon, ``Zur Lichtfortpflanzung nach der Relativit\"atstheorie'', Ann. Phys. (Leipzig), {\bf72} (1923) 421--456. 
doi: 10.1002/andp.19233772202

\bibitem{Unruh:1980}
  W.~G.~Unruh,
  ``Experimental black hole evaporation'',
  Phys.\ Rev.\ Lett.\  {\bf 46} (1981) 1351.
  %%CITATION = PRLTA,46,1351;%%
  
  \bibitem{LRR}
  C.~Barcel\'o, S.~Liberati and M.~Visser,
  ``Analogue gravity'',
  Living Rev.\ Rel.\  {\bf 8} (2005) 12,
   [Updated as Living Rev.\ Rel.\  {\bf 14} (2011) 3],
  [gr-qc/0505065].
  %%CITATION = GR-QC/0505065;%%
  
  \bibitem{unexpected}
  M.~Visser,
  ``Acoustic propagation in fluids: An Unexpected example of Lorentzian geometry'', unpublished, 
  gr-qc/9311028.
  %%CITATION = GR-QC/9311028;%%
  
  \bibitem{Visser:1997}
  M.~Visser,
  ``Acoustic black holes: Horizons, ergospheres, and Hawking radiation'',
  Class.\ Quant.\ Grav.\  {\bf 15} (1998) 1767
  [gr-qc/9712010].
  %%CITATION = GR-QC/9712010;%%
  
  \bibitem{Visser:2001}
  M.~Visser, C.~Barcel\'o and S.~Liberati,
  ``Analog models of and for gravity'',
  Gen.\ Rel.\ Grav.\  {\bf 34} (2002) 1719
  [gr-qc/0111111].
  %%CITATION = GR-QC/0111111;%%
  
 \bibitem{carmen}
M.~Visser and C.~Molina-Par\'is,
  ``Acoustic geometry for general relativistic barotropic irrotational fluid flow'',
  New J.\ Phys.\  {\bf 12} (2010) 095014
  [arXiv:1001.1310 [gr-qc]].
  %%CITATION = ARXIV:1001.1310;%%
  
 \bibitem{Prado}
  V.~Baccetti, P. Martin--Moruno, and M. Visser, 
  ``Gordon and Kerr--Schild ans\"atze in massive and bimetric gravity'',
  in preparation. 
  
  %---------------------------------------------------------------------------------------------------------------------------------------------

\bibitem{Tolman:1931}
Richard C. Tolman,
``On Thermodynamic Equilibrium in a Static Einstein Universe'',
Proc. Natl. Acad. Sci. USA {\bf17}  (1931)153--160.

\bibitem{Robertson:1933}  
H. P. Robertson,
``Relativistic Cosmology'',
Rev. Mod. Phys. {\bf5} (1933) 62--90.

\bibitem{Tolman:1939}
Richard C. Tolman,
``Static Solutions of Einstein's Field Equations for Spheres of Fluid'',
Phys. Rev. {\bf55} (1939) 364--373.

%---------------------------------------------------------------------------------------------------------------------------------------------

\bibitem{Maxwell1}
W.~D.~Niven, ed., \emph{The Scientific Papers of James Clerk Maxwell}, 1890. (Dover, New York, 2003), p. 76.  ISBN 0486495612, 9780486495613.
    
\bibitem{Maxwell2}
Anonymous,  ``Problems (3)'', The Cambridge and Dublin Mathematical Journal (Macmillan) {\bf8} (1853) 188.

\bibitem{Maxwell3}
Anonymous, ``Solutions of problems (prob. 3, vol. VIII. p. 188)'', The Cambridge and Dublin Mathematical Journal (Macmillan) {\bf9} (1854) 9--11.
  
%---------------------------------------------------------------------------------------------------------------------------------------------
  
\bibitem{Luneburg}
 R.~K.~L\"uneburg, \emph{Mathematical Theory of Optics}, (Brown University, Providence, Rhode Island, 1944), pp. 189--213.
 
 \bibitem{Leonhardt:2006}
  Ulf Leonhardt, 
  ``Optical Conformal Mapping",
  Science {\bf312} (2006) 1777--1780.  
  DOI:10.1126/science.1126493.
  
   \bibitem{Pendry:2006}
   J.~B.~Pendry, D.~Schurig, and D.~R.~Smith, 
   ``Controlling Electromagnetic Fields",
   Science {\bf312} (2006) 1780--1782. 
   DOI:10.1126/science.1125907.
  
  \bibitem{Leonhardt:2006b}
  Ulf Leonhardt and Thomas Philbin, 
  ``General relativity in electrical engineering",
  New Journal of Physics {\bf8} (2006) 247  [arXiv:cond-mat/0607418]. 
   DOI:10.1088/1367-2630/8/10/247.
   %%CITATION = COND-MAT/0607418;%%
  

 \bibitem{Leonhardt:2008}
   Ulf Leonhardt and Thomas Philbin, 
   ``Transformation Optics and the Geometry of Light'',
   Prog. Opt. {\bf53} (2009) 69--152 [arXiv:0805.4778].
   
%---------------------------------------------------------------------------------------------------------------------------------------------

\bibitem{metamaterials}
J.~Pendry,
  ``Optics: All smoke and metamaterials'',
  Nature {\bf 460} (2009) 579.
  %%CITATION = NATUA,460,579;%%

%---------------------------------------------------------------------------------------------------------------------------------------------

\bibitem{Kerr-slice}
 M.~Visser and S.~E.~C.~.Weinfurtner,
  ``Vortex geometry for the equatorial slice of the Kerr black hole'',
  Class.\ Quant.\ Grav.\  {\bf 22} (2005) 2493
  [gr-qc/0409014].
  %%CITATION = GR-QC/0409014;%%

%---------------------------------------------------------------------------------------------------------------------------------------------

\bibitem{QNM1}
E. Berti, V. Cardoso, J. P. S. Lemos, 
``Quasinormal modes and classical wave propagation in analogue black holes'', 
Physical Review D{\bf70} (2004) 124006; arXiv:gr-qc/0408099.

%---------------------------------------------------------------------------------------------------------------------------------------------

\bibitem{QNM2}
J. P. S. Lemos, ``Rotating acoustic holes: Quasinormal modes and tails,
super-resonance, and sonic bombs and plants in the draining bathtub'',
to appear in the Proceedings of the II Amazonian Symposium on Physics -- Analogue Models of Gravity.

%---------------------------------------------------------------------------------------------------------------------------------------------

  \bibitem{Stone}
  S.~E.~Perez Bergliaffa, K.~Hibberd, M.~Stone and M.~Visser,
  ``Wave equation for sound in fluids with vorticity'',
  Physica D {\bf 191} (2004) 121
  [cond-mat/0106255].
  %%CITATION = COND-MAT/0106255;%%
  
  %---------------------------------------------------------------------------------------------------------------------------------------------
  
  \bibitem{Kang}
   T.~Jacobson and G.~Kang,
  ``Conformal invariance of black hole temperature'',
  Class.\ Quant.\ Grav.\  {\bf 10} (1993) L201
  [gr-qc/9307002].
  %%CITATION = GR-QC/9307002;%%
  
  %-------------------------------------------------------------------
 \bibitem{Moncrief} %Moncrief
 V.~Moncrief, 
 ``Stability of stationary, spherical accretion onto a Schwarzschild black hole'', 
 Astrophys. J., {\bf 235} (1980) 1038--1046.
 
 
 %-------------------------------------------------------------------
 \bibitem{Vikman} %Vikman
  E.~Babichev, V.~Mukhanov and A.~Vikman,
  ``k-Essence, superluminal propagation, causality and emergent geometry'',
  JHEP {\bf 0802} (2008) 101
  [arXiv:0708.0561 [hep-th]].
  %%CITATION = ARXIV:0708.0561;%%
  
%---------------------------------------------------------------------------------------------------------------------------------------------

 \bibitem{Fagnocchi:2010}
  S.~Fagnocchi, S.~Finazzi, S.~Liberati, M.~Kormos and A.~Trombettoni,
  ``Relativistic Bose-Einstein Condensates: a New System for Analogue Models of Gravity'',
  New J.\ Phys.\  {\bf 12} (2010) 095012
  [arXiv:1001.1044 [gr-qc]].
  %%CITATION = ARXIV:1001.1044;%%

%---------------------------------------------------------------------------------------------------------------------------------------------



 \bibitem{Garay:1999}
  L.~J.~Garay, J.~R.~Anglin, J.~I.~Cirac and P.~Zoller,
  ``Black holes in Bose-Einstein condensates'',
  Phys.\ Rev.\ Lett.\  {\bf 85} (2000) 4643
  [gr-qc/0002015].
  %%CITATION = GR-QC/0002015;%%
  
  \bibitem{Barcelo:2000}
  C.~Barcel\'o, S.~Liberati and M.~Visser,
  ``Analog gravity from Bose-Einstein condensates'',
  Class.\ Quant.\ Grav.\  {\bf 18} (2001) 1137
  [gr-qc/0011026].
  %%CITATION = GR-QC/0011026;%%
  
  \bibitem{Barcelo:2001}
  C.~Barcel\'o, S.~Liberati and M.~Visser,
  ``Towards the observation of Hawking radiation in Bose-Einstein condensates'',
  Int.\ J.\ Mod.\ Phys.\ A {\bf 18} (2003) 3735
  [gr-qc/0110036].
  %%CITATION = GR-QC/0110036;%%
  
  \bibitem{Barcelo:2003}
  C.~Barcel\'o, S.~Liberati and M.~Visser,
  ``Probing semiclassical analog gravity in Bose-Einstein condensates with widely tunable interactions'',
  Phys.\ Rev.\ A {\bf 68} (2003) 053613
  [cond-mat/0307491].
  %%CITATION = COND-MAT/0307491;%%
    
  \bibitem{Lahav:2009}
  O.~Lahav, A.~Itah, A.~Blumkin, C.~Gordon and J.~Steinhauer,
  ``Realization of a sonic black hole analogue in a Bose-Einstein condensate'',
  Phys.\ Rev.\ Lett.\  {\bf 105} (2010) 240401
  [arXiv:0906.1337 [cond-mat.quant-gas]].
  %%CITATION = ARXIV:0906.1337;%%
  
 

   \bibitem{Jannes:2009}
  G.~Jannes,
  ``Emergent gravity: the BEC paradigm'',
  arXiv:0907.2839 [gr-qc].
  %%CITATION = ARXIV:0907.2839;%%
  
  \bibitem{Mayoral:2010}
  C.~Mayoral, A.~Recati, A.~Fabbri, R.~Parentani, R.~Balbinot and I.~Carusotto,
  ``Acoustic white holes in flowing atomic Bose-Einstein condensates'',
  New J.\ Phys.\  {\bf 13} (2011) 025007
  [arXiv:1009.6196 [cond-mat.quant-gas]].
  %%CITATION = ARXIV:1009.6196;%%

  
 %---------------------------------------------------------------------------------------------------------------------------------------------

\bibitem{Balbinot:Como}
  R. Balbinot, I. Carusotto, A. Fabbri, C. Mayoral, and A. Recati,
  ``Understanding Hawking radiation from simple BEC models'',
  these proceedings.

%---------------------------------------------------------------------------------------------------------------------------------------------

\bibitem{Visser:2005}
  M.~Visser and S.~Weinfurtner,
  ``Massive Klein-Gordon equation from a BEC-based analogue spacetime'',
  Phys.\ Rev.\ D {\bf 72} (2005) 044020
  [gr-qc/0506029].
  %%CITATION = GR-QC/0506029;%%
  
  \bibitem{Visser:2004}
  M.~Visser and S.~Weinfurtner,
  ``Massive phonon modes from a BEC-based analog model'',
  cond-mat/0409639.
  %%CITATION = COND-MAT/0409639;%%
  
 \bibitem{Liberati:2005}
  S.~Liberati, M.~Visser and S.~Weinfurtner,
  ``Naturalness in emergent spacetime'',
  Phys.\ Rev.\ Lett.\  {\bf 96} (2006) 151301
  [gr-qc/0512139].
  %%CITATION = GR-QC/0512139;%%

  
 %---------------------------------------------------------------------------------------------------------------------------------------------  
%---------------------------------------------------------------------------------------------------------------------------------------------


  
 \bibitem{Badulin}
  S.~I.~Badulin, K.~V.~Pokazayev, and A.~D.~Rozenberg, 
  ``A laboratory study of the transformation of regular gravity-capillary waves in inhomogeneous flows'', 
  Izv. Atmos. Ocean. Phys.  {\bf19} (1983) 782--787. % 19(10)?

    
  \bibitem{Silke&Bill}
  S.~Weinfurtner, E.~W.~Tedford, M.~C.~J.~Penrice, W.~G.~Unruh and G.~A.~Lawrence,
  ``Measurement of stimulated Hawking emission in an analogue system'',
  Phys.\ Rev.\ Lett.\  {\bf 106} (2011) 021302
  [arXiv:1008.1911 [gr-qc]].
  %%CITATION = ARXIV:1008.1911;%%
  
  \bibitem{Rousseaux:2007}
  G.~Rousseaux, C.~Mathis, P.~Maissa, T.~G.~Philbin and U.~Leonhardt,
  ``Observation of negative phase velocity waves in a water tank: A classical analogue to the Hawking effect?'',
  New J.\ Phys.\  {\bf 10} (2008) 053015
  [arXiv:0711.4767 [gr-qc]].
  %%CITATION = ARXIV:0711.4767;%%
  
  %---------------------------------------------------------------------------------------------------------------------------------------------
  
   

  \bibitem{toy}
   M.~Visser and S.~Weinfurtner,
  ``Analogue spacetimes: Toy models for quantum gravity'',
  PoS  {\bf QG-PH} (2007) 042
  [arXiv:0712.0427 [gr-qc]].
  %%CITATION = ARXIV:0712.0427;%%
  
  \bibitem{pedagogical}
  M.~Visser,
  ``Emergent rainbow spacetimes: Two pedagogical examples'',
  Time \& Matter 2007, Lake Bled, Slovenia,
  arXiv:0712.0810 [gr-qc].
  %%CITATION = ARXIV:0712.0810;%%
  
   \bibitem{Rousseaux:2010}
  G.~Rousseaux, P.~Maissa, C.~Mathis, P.~Coullet, T.~G.~Philbin and U.~Leonhardt,
  ``Horizon effects with surface waves on moving water'',
  New J.\ Phys.\  {\bf 12} (2010) 095018
  [arXiv:1004.5546 [gr-qc]].
  %%CITATION = ARXIV:1004.5546;%%
  
  \bibitem{Rousseaux:Como}
 Germain Rousseaux,
 ``The Basics of Water Waves Theory for Analogue Gravity'',
 these proceedings, arXiv:1203.3018v1 [physics.flu-dyn].

  
 %---------------------------------------------------------------------------------------------------------------------------------------------
  
     
 \bibitem{Boussinesq:1871}
  J. Boussinesq,  ``Th\'eorie de l'intumescence liquide, applel\'ee onde solitaire ou de translation, se propageant dans un canal rectangulaire'', Comptes Rendus de l'Academie des Sciences {\bf72} (1871) 755--759.
  
 \bibitem{Boussinesq:1872}
J. Boussinesq, ``Th\'eorie des ondes et des remous qui se propagent le long d'un canal rectangulaire horizontal, en communiquant au liquide contenu dans ce canal des vitesses sensiblement pareilles de la surface au fond'', Journal de Math\'ematiques Pures et Appliqu\'ees. Deuxi\`eme S\'erie {\bf17} (1872) 55--108.
  
  
  \bibitem{Madsen}
  P.~A.~Madsen and H.~A.~Schaffer, 
  ``Higher-order Boussinesq type equations for surface gravity waves: Derivation and analysis'', 
  Philosophical Transactions: Mathematical, Physical, and Engineering Sciences, {\bf356} (1998) 3123--3184. 
  See {\sf http://www.jtsor.org/stable/55084}
  
  %---------------------------------------------------------------------------------------------------------------------------------------------

  \bibitem{Schutzhold:2002}
  R.~Sch\"utzhold and W.~G.~Unruh,
  ``Gravity wave analogs of black holes'',
  Phys.\ Rev.\ D {\bf 66} (2002) 044019
  [gr-qc/0205099].
  %%CITATION = GR-QC/0205099;%%
  
  \bibitem{Unruh:Como}
  W.~G.~Unruh,
  ``Irrotational, two-dimensional Surface waves in fluids'',
  these proceedings, 
  arXiv:1205.6751 [gr-qc].
  %%CITATION = ARXIV:1205.6751;%%
  

  \bibitem{Chaline:Como}
  J.~Chaline, G.~Jannes, P.~Maissa and G.~Rousseaux,
  ``Some aspects of dispersive horizons: lessons from surface waves'',
  these proceedings, 
  arXiv:1203.2492 [physics.flu-dyn].
  %%CITATION = ARXIV:1203.2492;%%
  
  \bibitem{Carusotto:Como}
  I.~Carusotto and G.~Rousseaux,
  ``The Cerenkov effect revisited: From swimming ducks to zero modes in gravitational analogs'',
  these proceedings,
  arXiv:1202.3494 [physics.class-ph].
  %%CITATION = ARXIV:1202.3494;%%
  

  
%---------------------------------------------------------------------------------------------------------------------------------------------
  
  \bibitem{Jannes:2010}
  G.~Jannes, R.~Piquet, P.~Maissa, C.~Mathis and G.~Rousseaux,
  ``Experimental demonstration of the supersonic-subsonic bifurcation in the circular jump: A hydrodynamic white hole'',
  Phys.\ Rev.\ E {\bf 83} (2011) 056312
  [arXiv:1010.1701 [physics.flu-dyn]].
  %%CITATION = ARXIV:1010.1701;%%
  
  \bibitem{Volovik:2005}
  G.~E.~Volovik,
  ``The Hydraulic jump as a white hole'',
  JETP Lett.\  {\bf 82} (2005) 624
   [Pisma Zh.\ Eksp.\ Teor.\ Fiz.\  {\bf 82} (2005) 706]
  [physics/0508215].
  %%CITATION = PHYSICS/0508215;%%
  
%---------------------------------------------------------------------------------------------------------------------------------------------

  
  \bibitem{fibre}
  T.~G.~Philbin, C.~Kuklewicz, S.~Robertson, S.~Hill, F.~Konig, U.~Leonhardt,
  ``Fiber-optical analogue of the event horizon'',
  Science {\bf 319}, 1367-1370 (2008).
  [arXiv:0711.4796 [gr-qc]].
  
%---------------------------------------------------------------------------------------------------------------------------------------------

  \bibitem{glass1}
   F.~Belgiorno, S.~L.~Cacciatori, G.~Ortenzi, V.~G.~Sala, D.~Faccio,
  ``Quantum Radiation from Superluminal Refractive-Index Perturbations'',
  Phys.\ Rev.\ Lett.\  {\bf 104}, 140403 (2010).
  [arXiv:0910.3508 [quant-ph]].
 
   \bibitem{glass2}
  F.~Belgiorno, S.~L.~Cacciatori, G.~Ortenzi, L.~Rizzi, V.~Gorini, D.~Faccio,
  ``Dielectric black holes induced by a refractive index perturbation and the Hawking effect'',
  Phys.\ Rev.\  {\bf D83}, 024015 (2011).
  [arXiv:1003.4150 [quant-ph]].
 
   \bibitem{glass3}
 S.~L.~Cacciatori, F.~Belgiorno, V.~Gorini, G.~Ortenzi, L.~Rizzi, V.~G.~Sala, D.~Faccio,
  ``Space-time geometries and light trapping in travelling refractive index perturbations'',
  New J.\ Phys.\  {\bf 12}, 095021 (2010).
  [arXiv:1006.1097 [physics.optics]].

 %---------------------------------------------------------------------------------------------------------------------------------------------
    
   \bibitem{Belgiorno:2010}
  F.~Belgiorno, S.~L.~Cacciatori, M.~Clerici, V.~Gorini, G.~Ortenzi, L.~Rizzi, E.~Rubino and V.~G.~Sala {\it et al.},
  ``Hawking radiation from ultrashort laser pulse filaments'',
  arXiv:1009.4634 [gr-qc].
  %%CITATION = ARXIV:1009.4634;%%
  
  \bibitem{Schutzhold:2010}
  R.~Sch\"utzhold and W.~G.~Unruh,
  ``Comment on: Hawking Radiation from Ultrashort Laser Pulse Filaments'',
  Phys.\ Rev.\ Lett.\  {\bf 107} (2011) 149401
  [arXiv:1012.2686 [quant-ph]].
  %%CITATION = ARXIV:1012.2686;%%
  
  \bibitem{Belgiorno:2010b}
  F.~Belgiorno, S.~L.~Cacciatori, M.~Clerici, V.~Gorini, G.~Ortenzi, L.~Rizzi, E.~Rubino and V.~G.~Sala {\it et al.},
  ``Reply to Comment on: Hawking radiation from ultrashort laser pulse filaments'',
  Phys.\ Rev.\ Lett.\  {\bf 107} (2011) 149402
  [arXiv:1012.5062 [quant-ph]].
  %%CITATION = ARXIV:1012.5062;%%
  
  \bibitem{Prain}
  S.~Liberati, A.~Prain and M.~Visser,
  ``Quantum vacuum radiation in optical glass'',
  Phys. Rev. D (in press),
  arXiv:1111.0214 [gr-qc].
  %%CITATION = ARXIV:1111.0214;%%
  
  %---------------------------------------------------------------------------------------------------------------------------------------------
  
  \bibitem{pulse:2005}
R.~Sch\"utzhold and W.~G.~Unruh, 
``Hawking Radiation in an Electromagnetic Waveguide?",
Phys.\ Rev.\ Lett.\ {\bf 95}, 031301 (2005).

\bibitem{Cortijo:2006}
  A.~Cortijo and M.~A.~H.~Vozmediano,
  ``Effects of topological defects and local curvature on the electronic properties of planar graphene'',
  Nucl.\ Phys.\ B {\bf 763} (2007) 293
   [Nucl.\ Phys.\ B {\bf 807} (2009) 659]
  [cond-mat/0612374].
  %%CITATION = COND-MAT/0612374;%%
  
  \bibitem{Vozmediano:2010}
  M.~A.~H.~Vozmediano, M.~I.~Katsnelson and F.~Guinea,
  ``Gauge fields in graphene'',
  Phys.\ Rept.\  {\bf 496} (2010) 109.
  %%CITATION = PRPLC,496,109;%%
  
 \bibitem{Leonhardt:2000}
  U.~Leonhardt and P.~Piwnicki,
  ``Relativistic effects of light in moving media with extremely low group velocity'',
  Phys.\ Rev.\ Lett.\  {\bf 84} (2000) 822.
  %%CITATION = PRLTA,84,822;%%
  
  \bibitem{Visser:2000}
  M.~Visser,
  ``Comment on: Relativistic effects of light in moving media with extremely low group velocity'',
  Phys.\ Rev.\ Lett.\  {\bf 85} (2000) 5252
  [gr-qc/0002011].
  %%CITATION = GR-QC/0002011;%%
  
  \bibitem{Leonhardt:2000b}
  U.~Leonhardt and P.~Piwnicki,
  ``Reply to the Comment on `Relativistic effects of light in moving media with extremely low group velocity' by M. Visser'',
  Phys.\ Rev.\ Lett.\  {\bf 85} (2000) 5253
  [gr-qc/0003016].
  %%CITATION = GR-QC/0003016;%%

\bibitem{Leonhardt:2000c}
  U.~Leonhardt,
  ``Space-time geometry of quantum dielectrics'',
  physics/0001064.
  %%CITATION = PHYSICS/0001064;%%

\bibitem{Leonhardt:2001}
  U.~Leonhardt,
  ``A Primer to slow light'',
  gr-qc/0108085.
  %%CITATION = GR-QC/0108085;%%
  
   \bibitem{Volovik:2000}
  G.~E.~Volovik,
  ``Superfluid analogies of cosmological phenomena'',
  Phys.\ Rept.\  {\bf 351} (2001) 195
  [gr-qc/0005091].
  %%CITATION = GR-QC/0005091;%%
  
  \bibitem{Jacobson:1998}
  T.~A.~Jacobson and G.~E.~Volovik,
  ``Event horizons and ergoregions in He-3'',
  Phys.\ Rev.\ D {\bf 58} (1998) 064021.
  %%CITATION = PHRVA,D58,064021;%%
  
%---------------------------------------------------------------------------------------------------------------------------------------------
  
 
 \bibitem{Giovanazzi:2004}
  S.~Giovanazzi,
  ``Hawking radiation in sonic black holes'',
  Phys.\ Rev.\ Lett.\  {\bf 94} (2005) 061302
  [physics/0411064].
  %%CITATION = PHYSICS/0411064;%%

\bibitem{Giovanazzi:2011}
  S.~Giovanazzi,
  ``Entanglement Entropy and Mutual Information Production Rates in Acoustic Black Holes'',
  Phys.\ Rev.\ Lett.\  {\bf 106} (2011) 011302
  [arXiv:1101.3272 [cond-mat.other]].
  %%CITATION = ARXIV:1101.3272;%%
  
  \bibitem{Horstmann:2009}
  B.~Horstmann, B.~Reznik, S.~Fagnocchi and J.~I.~Cirac,
  ``Hawking Radiation from an Acoustic Black Hole on an Ion Ring'',
  Phys.\ Rev.\ Lett.\  {\bf 104} (2010) 250403
  [arXiv:0904.4801 [quant-ph]].
  %%CITATION = ARXIV:0904.4801;%%
  
  
  
%--------------------------------------------------------------------------------------------------------------------------------------------- 

 
\bibitem{Liberati:2012}
  S.~Liberati,
  ``Lorentz breaking: Effective Field Theory and observational tests'',
  these proceedings, 
  arXiv:1203.4105 [gr-qc].
  %%CITATION = ARXIV:1203.4105;%%
  
 %---------------------------------------------------------------------------------------------------------------------------------------------
 
 \bibitem{Volovik:2011kg}
  G.~E.~Volovik,
  ``Topology of quantum vacuum'',
  these proceedings,
  arXiv:1111.4627 [hep-ph].
  %%CITATION = ARXIV:1111.4627;%%
 %---------------------------------------------------------------------------------------------------------------------------------------------
 
 \bibitem{Jannes:2011em}
  G.~Jannes and G.~E.~Volovik,
  ``The cosmological constant: A lesson from the effective gravity of topological Weyl media'',
  arXiv:1108.5086 [gr-qc].
  %%CITATION = ARXIV:1108.5086;%%
 %---------------------------------------------------------------------------------------------------------------------------------------------
 
 \bibitem{Finazzi:2012wz}
  S.~Finazzi, S.~Liberati and L.~Sindoni,
  ``The analogue cosmological constant in Bose-Einstein condensates: a lesson for quantum gravity'',
  arXiv:1204.3039 [gr-qc],
  to appear in the Proceedings of the II Amazonian Symposium on Physics -- Analogue Models of Gravity.
  %%CITATION = ARXIV:1204.3039;%%
 
 %---------------------------------------------------------------------------------------------------------------------------------------------
 %---------------------------------------------------------------------------------------------------------------------------------------------
 %---------------------------------------------------------------------------------------------------------------------------------------------
 %---------------------------------------------------------------------------------------------------------------------------------------------
 %---------------------------------------------------------------------------------------------------------------------------------------------
 %---------------------------------------------------------------------------------------------------------------------------------------------
 
  %--------------------------------------------------------------------------------------------------------------------------------------------- 
%--------------------------------------------------------------------------------------------------------------------------------------------- 
%--------------------------------------------------------------------------------------------------------------------------------------------- 
\end{thebibliography}
\end{document}